\documentstyle[aas2pp4]{article}

\begin{document}

\title{Quasars as Cosmological Probes:\\
The Ionizing Continuum, Gas Metallicity and the $W_\lambda$-L Relation}

\author{Kirk Korista,}
\affil{Department of Physics, Western Michigan University,
Kalamazoo, MI  49008}

\author{ Jack Baldwin\altaffilmark{1}}
\affil{Cerro Tololo Interamerican Observatory, National Optical
Astronomy Observatories, Tucson, AZ 85726}

\and 

\author{Gary Ferland}
\affil{Department of Physics and Astronomy, University of Kentucky,
Lexington, KY 40506}

\altaffiltext{1}{CTIO is operated by AURA, Inc.\ under contract to the
National Science Foundation.}

\date{\today}

\begin{abstract}

Using a realistic model for line emission from the broad emission line
regions of quasars, we are able to reproduce the previously observed
correlations of emission-line ratios with the shape of the spectral
energy distribution (SED). In agreement with previous studies, we find
that the primary driving force behind the Baldwin Effect ($W_\lambda\/
\propto L^{\beta}$, $\beta < 0$) is a global change in the SED with
quasar luminosity, in that more luminous quasars must have
characteristically softer ionizing continua. This is completely
consistent with observations that show (1) a correlation between
$L_{uv}$ and $\alpha_{ox}$ and $\alpha_{uvx}$, (2) correlations of SED
shape sensitive line ratios with $\alpha_{ox}$, $\alpha_{uvx}$, and
$L_{uv}$, and (3) correlations between line equivalent widths and
$\alpha_{ox}$, $\alpha_{uvx}$, and $L_{uv}$.  However, to explain the
complete lack of a correlation in the $W_\lambda$(\ion{N}{5}) --
$L_{uv}$ diagram we propose that the more luminous quasars have
characteristically larger gas metallicities ($Z$). As a secondary
element, nitrogen's rapidly increasing abundance with increasing $Z$
compensates for the losses in $W_\lambda$(\ion{N}{5}) emitted by gas
illuminated by softer continua in higher luminosity quasars. A
characteristic relationship between $Z$ and $L$ has an impact on the
$W_\lambda$ -- $L_{uv}$ relations for other lines as well.  For a {\em
fixed} SED, an increasing gas metallicity reduces the $W_\lambda$ of
the stronger metal lines (the gas cools) and that of Ly$\alpha$ and
especially \ion{He}{2} (because of the increasing metal opacity), while
the weaker lines (e.g., \ion{C}{3}] 1909) generally respond
positively.  The interplay between the effects of a changing SED and
$Z$ with $L$ results in the observed luminosity dependent spectral
variations.  All of the resulting dependences on $L_{uv}$ are within
the range of the observed slopes.

\end{abstract}


\section{Introduction}

Since their discovery as the most distant and luminous discrete objects
in the Universe, it has been hoped that somehow quasars could be used
as ``cosmological candles'' to measure the expansion parameters ($q_o$
and $H_o$) of the universe. With the discovery of the Baldwin Effect
(Baldwin 1977), whereby the emission line equivalent widths scale
inversely with the quasar luminosity, that hope is half-realized.
Unfortunately, the very large scatter in these correlations (Baldwin,
Wampler \& Gaskell 1989; Kinney, Rivolo \& Koratkar 1990; Osmer, Porter
\& Green 1994) have curtailed their usefulness. What remains is to
understand the mechanism of the formation of quasar emission lines well
enough so that this scatter can somehow be calibrated in terms of
observable parameters, and then removed.

A second major goal of studies of distant quasars is to use them as
probes of the first generations of nucleosynthesis in the young
Universe. There is evidence of wide variations in the atomic abundances
in the BLRs of different quasars, and that the spectroscopic effects of
this can be measured out to very large redshifts (Hamann \& Ferland
1992; Hamann \& Ferland 1993: HF93; Ferland et al 1996). Again, full
interpretation requires a proper understanding of the formation of
quasar emission lines.

There is mounting evidence that the line strengths strongly depend on
the spectral energy distribution (SED) in the UV -- X-Ray range.
(Netzer, Laor, and Gondhalekar 1992, Zheng et al. 1995, Green 1996,
Wang et al.  1998). In addition, many studies have found a correlation
between $\alpha_{ox}$ and luminosity (cf.  Zamorani et al.  1981;
Wilkes et al. 1994; but see also LaFranca et al 1995 and Avni et al
1995 for discussion of conflicting results).  This link through
$\alpha_{ox}$ offers a promising explanation of the observed
$W_\lambda$--$L_{uv}$ relations (Wang et al 1998; Green 1998).  Here we
investigate to what extent we can reproduce these observed effects
using realistic photoionization models of the BLR, and study the degree
to which the additional parameter of metallicity must also be taken
into account.

\section{LOC Models}

We describe the BLR in terms of the ``Locally Optimally-emitting
Clouds'' (LOC) model, discussed in Baldwin et al. (1995). This model
assumes that the many individual gas clouds which make up the BLR have
a wide range of internal densities and sizes and occur over a wide
range in distance from the central continuum source. Under these
conditions the emitted spectrum is controlled by powerful selection
effects introduced by the atomic physics and basic radiative transfer
effects (see also Korista et al. 1997), and the typical observed quasar
spectrum is naturally produced.

This picture is supported by the line-continuum reverberation studies,
showing that gas having a range of densities must exist over a wide
range of radii. (Ferland et al.\ 1992; Peterson 1993).  The spectrum
predicted by the LOC depends on global integrations, over gas density
at a particular location, and over radius.  Results depend weakly on
the density distribution near those distributions that produce typical
quasar spectra, and only somewhat on the radial distribution.  This can
be contrasted to single cloud models, which often are described by an
ionization parameter (Davidson \& Netzer 1979) and whose predicted
spectrum has a powerful dependence on this parameter. The LOC models
fit the observed spectrum better with fewer free parameters than do
single cloud models. But more importantly, the LOC approach is a more
physical model because unless the BLR clouds have a remarkably
restricted range of properties, we will in fact observe the optimally
emitting clouds for most lines. The emission line spectrum from clouds
distributed in gas density and radius is much less sensitive to changes
in the gas abundances and SED than is that emitted by a single cloud of
a fixed gas density and ionization parameter. Thus analyses that use a
single cloud to investigate the expected spectral variations with
changes in gas abundances or SED may derive misleading results.

We have generated extensive grids of photoionization models similar to
those shown in Korista et al. (1997), and integrated over the cloud
properties to obtain predicted LOC spectra.  We did this for a wide
range of SED and metallicities, with the goal of finding the dependence
of the final LOC spectra on these two parameters. To the extent that
the geometry is independent of  luminosity, this comparison will not be
affected at all by the assumed cloud distributions, for which we have
just used the standard LOC parameters: $f(r) \propto r^{-1}$, an outer
radius corresponding to a hydrogen ionizing photon flux of $10^{18}$
s$^{-1}$cm$^{-2}$, and $g(n_H) \propto n^{-1}$, integrated from $\log
n_H = 8$ to $\log n_H = 12$ cm$^{-3}$ (see Baldwin et al. 1995 for
definitions of $f(r)$ and $g(n_H)$). For simplicity we took the
ionizing continuum shape to be a single power law, $f_{\nu} \propto
\nu^{\alpha}$, and varied the spectral index over the range $-2 <
\alpha < -1$, normalizing each SED to have the same number of photons/s
below 912~\AA. Our prescription for abundances is given in Table~1,
where ``metals'' refers to all elements except H, He and N. These
variations with increasing $Z$ (= metals/H relative to the solar
abundance ratio) are similar to the HF93 galactic chemical enrichment
models whose major feature is that nitrogen is built up as a delayed
secondary element, at a rate N/H $= b(Z) Z^2$, with $b(Z) < 1$. Here we
have chosen $b(Z) = b = 0.5$, consistent with the rapid star formation
enrichment models presented in HF93. This particular choice of $b(Z)$
does not affect the structure of the cloud or the strengths of any
non-nitrogen lines in any significant manner. The helium abundance is
scaled gently with $Z$ as in HF93, however, for simplicity we have
chosen to scale all metals (excluding N) with $Z$ from solar
abundances.

Figure~1 shows the results for some of the stronger UV lines. The upper
panels show predicted equivalent widths as a function of $\alpha$ and
$Z$, while the bottom panels show corresponding intensity ratios. As
stressed by HF93, the strength of \ion{N}{5} $\lambda$1240 relative to
either \ion{C}{4} $\lambda$1549 or \ion{He}{2} $\lambda$1640 is a good
metallicity indicator, as can be seen in the lower-left panel. The
lower-right panel shows these same ratios for $Z = 1$ and various
continuum shapes.  The \ion{N}{5}/\ion{He}{2} and \ion{N}{5}/\ion{C}{4}
line ratios have little dependence on $\alpha$.  Although the
\ion{N}{5}/\ion{He}{2} ratio does become sensitive to the continuum
shape if a UV bump is introduced (factor of $\approx 2$ variation; see
also Ferland et al.\ 1996), these variations are relatively small
compared to the order of magnitude range of observed ratios. The
\ion{N}{5}/\ion{C}{4} ratio is virtually independent ($\pm 0.05$ dex)
of any reasonable quasar continuum. Conversely, the ratios
Ly$\alpha$/\ion{C}{4} and Ly$\alpha$/\ion{O}{6} $\lambda$1034 are not
affected much at all by changes in the metallicity but depend strongly
on the continuum shape, showing the steepest dependence on $\alpha$ of
any of the ratios of strong lines which we examined.

There are several reasons for these dependencies. The \ion{O}{6} and
\ion{C}{4} lines have ionization potentials much higher than hydrogen.
The abundance of these ions relative to hydrogen is sensitive to the
form of the continuum between 13.6 eV and 100 eV.  This was the basis
of the earlier investigations by Zheng et al. (1995), Green (1996) and
others. The intensities of these lines depend only weakly on the
metallicity because of the strong thermostat effect introduced by such
strong cooling lines.  As the abundance of C or O goes up the gas cools
more effectively and so the temperature falls as the stronger lines
maintain or even diminish in their intensities.  Nitrogen lines do
introduce a strong metallicity dependence since N is initially a rare
element, and its abundance relative to hydrogen goes up as the square
of the overall metallicity.  This large increase in abundance does
allow the \ion{N}{5} to grow stronger as it takes on more of the
cooling. The equivalent width of \ion{He}{2} $\lambda$1640 diminishes
with increasing $Z$ as the metals become increasingly important sources
of opacity.  To a lesser degree this is also true of Ly$\alpha$
$\lambda$1216.

Ferland et al.\ (1992) and Shields, Ferland, \& Peterson (1995) pointed
out that Ly$\alpha$ may be significantly contaminated with other
emission (including \ion{C}{3} $\lambda$1176, \ion{S}{3}
$\lambda$1190, \ion{S}{5}] $\lambda$1198, \ion{Si}{3} $\lambda$1207,
and especially \ion{He}{2} $\lambda$1216 and \ion{O}{5}]
$\lambda$1218). However, the model predictions find that these are
contaminants at the level of 8\% -- 15\% the intensity of Ly$\alpha$
over the full range in SED and $Z$ shown in Figure~1.  Our models
predict a similar level of contamination to \ion{O}{6} $\lambda$1034
due to Ly$\beta$ $\lambda$1025. Neither set of contaminants is
important to the results presented here.

In addition to the results shown in Figure~1, we also examined the
behavior of the equivalent widths of the $\lambda 1400$ (sum of
\ion{Si}{4} $\lambda 1397$, \ion{O}{4}] $\lambda 1402$, \ion{S}{4}]
$\lambda 1405$) and $\lambda 1900$ (sum of \ion{C}{3}] $\lambda 1909$,
\ion{Si}{3}] $\lambda 1892$, \ion{Al}{3} $\lambda 1860$) blends and of
\ion{Mg}{2} $\lambda 2800$, and of all intensity ratios between the
different lines. Note that the predicted strength of the $\lambda 1900$
blend does not include emission from \ion{Fe}{3} UV~34, a major
contributor to this blend in some quasars (Baldwin et al.\ 1996; Laor
et al.\ 1997).  The four intensity ratios shown in Figures 1c and 1d
were chosen because they had the strongest dependence on either
$\alpha$ or $Z$, and therefore will give the best leverage for
measuring these underlying parameters.

\section{Comparison with Observations}

It is possible to directly test the model results with respect to
$\alpha$. The work of Zheng et al (1995) and Wang et al (1998) shows
that the intensity ratios \ion{O}{6}/Ly$\alpha$ and
\ion{C}{4}/Ly$\alpha$ directly correlate with an observed indicator of
the shape of the ionizing continuum, $\alpha_{ox}$. Wang et al. show
that their \ion{C}{4}/Ly$\alpha$ data set can be fitted by single-cloud
models, provided they choose the correct ionization parameter and gas
density. Figure~2 shows that the results from the LOC simulations, with
$Z = 1$, also generally reproduce the observed trends in these
intensity ratios, but without requiring adjustable parameters.

Many studies have now confirmed the existence of $W_\lambda$--$L_{uv}$
correlations for most of the strong emission lines in quasar spectra.
These results are summarized in columns 4 and 6 of Table~2, which list
the observed slopes and intercepts of these correlations, taken from
the references cited in column 7. These observed relationships are such
that the strongest effect is seen in \ion{O}{6} followed by \ion{C}{4},
\ion{He}{2}, and then Ly$\alpha$.  Figure~1b shows that the same
sequence results from the LOC models, on the assumption that $\alpha$
increases (spectrum becomes softer) with increasing luminosity.  This
lends credence to the idea that the Baldwin Effect is at least in part
due to a systematic change in the incident continuum shape with
luminosity.

However, there is a glaring discrepancy between the predicted trends in
Figures 1b and 1d and the observed behavior of the \ion{N}{5}
$\lambda$1240 line. The LOC models (and for that matter any
single-cloud model) predict that $W_\lambda$(\ion{N}{5}) should vary as
strongly with $\alpha$ (and therefore with luminosity) as does
$W_\lambda$(\ion{C}{4}), whereas the observations show almost no
correlation with $L_{uv}$. Equivalently, \ion{N}{5}/\ion{C}{4} should
be nearly independent of $L_{uv}$, but the work of HF93 shows that
higher values of \ion{N}{5}/\ion{C}{4} are systematically found at
higher redshifts and/or higher $L$. Figure~6 of Osmer et al. (1994)
clearly shows the near absence of an $W_\lambda$(\ion{N}{5})--$L_{uv}$
relation, and the correspondingly strong dependence of
\ion{N}{5}/\ion{C}{4} on luminosity, for high-redshift quasars.  This
must mean that continuum shape is not the only parameter.

\section{Metallicity as a second parameter}

Figures~1a, 1b suggest the following solution: while most lines'
$W_\lambda$ diminish with increasing luminosity due to a systematic
softening of the SED, a systematic increase in the gas metallicity $Z$
(and $\rm{N/H} \propto Z^2$) with quasar luminosity could, in the case
of \ion{N}{5}, compensate for this loss and result in an $W_\lambda$
that is roughly constant with the quasar luminosity. Note that
$W_\lambda$(\ion{N}{5}) decreases by roughly 0.8 dex over the range of
$1.2 < \alpha < 1.8$, while it increases by virtually the same amount
over the range of $1 < Z < 10$. The range in $\alpha$ was chosen to
represent the observed range for quasars (Zamorani et al. 1981; Wilkes
et al. 1994).  The range in $Z$ is within the range which might be at
least reasonable according to population synthesis models (HF93), and a
detailed study by Ferland et al (1996) shows that in some quasars $Z$
must be greater than at least 5. Thus for the observed range of SEDs,
there is a reasonable range in $\rm{N/H} \propto Z^2$ that can offset
the expected change in $W_\lambda$(\ion{N}{5}), provided that both SED
and $Z$ depend on $L$.

The models do not tell us how to map $\alpha$ into $L$. For the present
study, we simply adopt the empirical $\alpha_{ox} - L$ relation found
by  Wilkes et al.\ (1994); thus we adopt their choice of $H_o =
50$~$\rm{ km~s^{-1}~Mpc^{-1} }$ and $q_o = 0$. This gives the effective
2500~\AA\/ to 2~keV continuum slope ($\alpha_{ox}$); we made a slight
adjustment in its logarithmic zero point to reference the continuum
luminosity at 1550~\AA\/. Since the line equivalent widths and
continuum-shape sensitive line ratios are observed to correlate with
$\alpha_{ox}$, the latter must be strongly correlated with the true
measure of the ionizing continuum's hardness.  Wang et al.\ (1998)
defined an $\alpha_{uvx}$, a measure of the ratio of fluxes at
1350~\AA\/ and 1~keV, closer to the ionizing photons, and reaffirmed
this assertion in a large sample of quasars.

As an initial guess at the $L-Z$ relation, we assumed a metallicity $Z
\approx 1$ for the lowest luminosity sources and $Z \approx 10$ for the
highest. This is consistent with the slopes of the
\ion{N}{5}/\ion{C}{4} and \ion{N}{5}/\ion{He}{2} dependences on
$L_{uv}$ given by HF93 (their Fig. 7), when mapped back to metallicity
$Z$ using our Fig 1a. Since $L$ and $\alpha_{ox}$ are also correlated,
we simply assigned $Z = 1$ to $\alpha_{ox} = 1.2$, $Z = 10$ to
$\alpha_{ox} = 1.8$, and a power law in $\alpha_{ox}$ and $Z$ in
between. Then using the relationship between $L_{\nu}(1550)$ and
$\alpha_{ox}$, this example results in the simple power law relation $
\log Z = 0.183 \log (L_{\nu}(1550)/10^{30}) + 0.477$.  The range in
$\alpha_{ox}$ and $Z$ results in a range in $L_{\nu}(1550)$ of
$10^{27.40}$ to $10^{32.85}$ $\rm{ergs~s^{-1}~Hz^{-1}}$.

In the spectral simulations, we assumed a simple power law continuum
SED and so $\alpha_{ox}$ is identical to $\alpha$: 1.2, 1.4, 1.6, 1.8.
In accordance with the above expression each power law SED was then
assigned to its respective metallicity $Z$: 1, 2, 5, 10. Four separate
grids of cloud emission were then computed and integrated, as described
above. The resulting predicted emission line/continuum relations and
their observed counterparts are given in Table~2. In each case the
predicted relationships were fitted with a straight line, usually a
good approximation. The slopes $\beta$ and intercepts $c$ are listed in
the table, where the fit is to $y = c + \beta x$. In four cases, a
significant break in the slope of the predicted relation forced a
double-power law fit, and both relations are given. The simulated
$W_\lambda$ for each emission line is referenced to the continuum at
1215~\AA\/ and a global covering fraction of 0.5 is assumed. These
affect only the intercepts, not the slopes. Finally, we point out that
relationships involving $L_{\nu}(1550)$ in Table~2 are not strictly
comparable, since different investigators (column 7) assumed different
values of the deceleration parameter $q_o$ (all assumed  $H_o =
50$~$\rm{ km~s^{-1}~Mpc^{-1} }$). When known, the assumed values of
$q_o$ are listed in the footnotes of Table~2.  The effect of different
choices of $q_o$ is non-uniform in the $W_\lambda$--$L_{uv}$ plane, and
its effect on the $W_\lambda$--$L_{uv}$ slope can be comparable to the
typically quoted uncertainty in the slope's measurement of $\pm 0.04$.

Given the simplicity of our assumptions, the predicted relationships
match the observed ones surprisingly well. The general lack of a
$W_\lambda$--$L_{uv}$ relationship in \ion{N}{5} and the trends in the
different slopes for different lines are reproduced.  We note that
while the overall behavior of the \ion{N}{5}/\ion{He}{2} vs.
\ion{N}{5}/\ion{C}{4} relationship is matched very well by our models,
we do not reproduce the highest observed values.  The primary reason
for this is that our models do not predict a large enough
$W_\lambda$(\ion{N}{5}) at a given $L_{uv}$. Of the lines studied here,
this line will be by far the most sensitive to the choice of abundances
(see Figure~1b). Under our hypothesis, our simulations also predict
that the line ratios Ly$\alpha$/\ion{C}{4} and $\lambda$1900/\ion{C}{4}
should increase with increasing $L_{uv}$, as is observed (Osmer et
al.\ 1994).  This is simply a result of differing slopes of their
respective $W_\lambda$--$L_{uv}$ relations. Many investigators have
suggested this change in the spectrum to be due to a shift in
ionization parameter to lower values at larger $L_{uv}$ (e.g.,
Mushotzky \& Ferland 1984). However, we would argue that this change in
the line spectrum is due to a characteristic shift to softer incident
continua and higher $Z$ with increasing $L_{uv}$. Of these two effects,
the metallicity has the largest impact on the $\lambda$1900/\ion{C}{4}
ratio. Observationally, the challenge will be to tighten these and
other key global relationships. Table~2 shows that these stronger lines
of quasars do carry sufficient information to constrain the proposed
evolution of the SED and $Z$ with $L$.

\section{Summary and Discussion}

We have shown here that including a $Z$--$L$ effect in addition to a
SED--$L$ relation allows us to simultaneously reproduce the observed
correlations with $L$ of the equivalent widths and intensity ratios of
all of the strong UV lines, particularly \ion{N}{5}.  As a secondary
element, nitrogen's rapidly increasing abundance with increasing $Z$
compensates for the losses in $W_\lambda$(\ion{N}{5}) emitted by gas
illuminated by softer continua in higher luminosity quasars. For a {\em
fixed} SED, an increasing gas metallicity reduces the $W_\lambda$ of
the stronger lines of other heavy elements (the gas cools) and that of
Ly$\alpha$ and especially \ion{He}{2} (because of the increasing metal
opacity), while the weaker metal lines (e.g., intercombination lines
such as \ion{C}{3}] 1909) generally respond positively. Weak lines such
as \ion{O}{4}] $\lambda$1402, formed in the He$^{++}$ zone, suffer from
the significant drop in temperature and their intensities remain
roughly constant with increasing $Z$.  The interplay between the
effects of a changing SED and $Z$ with $L$ results in the observed
luminosity dependent spectral variations.

However, the photoionization models only predict relations with
$\alpha$ and $Z$. To get to relations with $L$ we have had to adopt
both an $\alpha$--$L$ relation, and a $Z$--$L$ relation. These two
relations or at least their combined effect must still be calibrated
empirically, from comparing emission line parameters to continuum
fluxes, before the Baldwin Effect can be used as a real luminosity
indicator for cosmological tests.

Why should we expect a global relationship between $Z$ and quasar
luminosity? HF93 and Ferland et al.\ (1996) showed that the average
\ion{N}{5}/\ion{C}{4} and \ion{N}{5}/\ion{He}{2} ratios increase
systematically with redshift and/or luminosity. They attributed this
change in line ratios to an increase in metallicity, as we do here.
HF93 proposed that the luminosity dependence of metallicity is really a
mass-metallicity relation.  Because of their deeper potential wells,
higher-mass galaxies are better able to retain gas deposited during
stellar evolution than are lower-mass galaxies, so they have a more
rapid buildup of heavy elements in their nuclear regions. Coupling this
scenario with observations showing that the most luminous quasars tend
to reside in the most massive galaxies (McLeod \& Rieke 1995a, 1995b;
Bahcall et al.\ 1997) leads to a natural explanation for a global
relationship between quasar luminosity and gas metallicity. To the
extent that luminosity evolution affects individual quasars, the
redshift-metallicity dependence would also require a steady decrease in
metallicity with time in individual objects. HF93 suggested that this
was due to dilution by lower-$Z$ gas falling into the nuclear regions
throughout the lives of these galaxies.

In addition, the simplest accretion disk models predict that more
luminous quasars should in general emit softer continua (Netzer et
al.\ 1992), though the continuum emission mechanism is far from
understood. Together, these form the basis of the proposed global
$L$--SED--$Z$ relationship.

Can the noise in the Baldwin Effect be eliminated, much as an observer
corrects for reddening?  Metallicity should scale with the mass of the
host galaxy and not necessarily the quasar luminosity, thus introducing
noise.  The importance of this effect can be estimated from Figure~3,
which compares the observations of Osmer et al. (1994) to the range
covered by our models on the $W_\lambda$(\ion{C}{4})--$L$ diagram as
metallicity is varied from half solar to ten times solar. Just making a
correction on the basis of the \ion{N}{5}/\ion{C}{4} ratio may remove
much of the scatter at high redshifts; this is something that can be
tested empirically. There are likely to be additional sources of
scatter as well: a spread in $\alpha$ at given $L$; the strength of any
additional EUV bump above a simple power-law continuum; variations in
the cloud distributions; line and continuum emission anisotropies;
variability (the effects of line--continuum lags and the intrinsic
Baldwin Effect; Kinney, Rivolo, \& Koratkar 1990; Pogge \& Peterson
1992). Our aim is to use the more accurate BLR model described here
together with the measurements of critical line and continuum strengths
to distinguish between as many of these effects as possible and guide
us in their empirical calibration.  Even a factor of two reduction in
the scatter would be an important gain in our ability to use these
diagrams for cosmology.

\acknowledgements{We are grateful for the careful reading and helpful
suggestions of the referee, Joe Shields. KTK and GJF acknowledge
support from NASA grant NAG-3223 and NSF grant AST 96-17083.}

\clearpage

\figcaption[fig1.eps]{ 
Equivalent widths for full geometric source coverage and intensity
ratios from LOC models. (a) $W_\lambda$ vs. log $Z$, for $\alpha = 1.4$;
(b) log $W_\lambda$ vs. $\alpha$, for log $Z$ = 0; (c) log ratio vs. log
$Z$, for $\alpha = 1.4$; (d) log ratio vs. $\alpha$, for log $Z$ = 0.}

\figcaption[fig2.eps]{ 
Comparison of LOC results for $Z = 1$ to observed \ion{C}{4}/Ly$\alpha$
(Wang et al.\ 1998), \ion{O}{6}/Ly$\alpha$ (Zheng et al.\ 1995). }

\figcaption[fig3.eps]{
$W_\lambda$(\ion{C}{4})--$L_{uv}$ relation calculated for the lowest
and highest metallicity $Z$ considered in our models with $\alpha =
-1.4$ (solid lines), and for the model in which both $Z$ and the SED
are varied together (dashed line), shown overplotted on the
observational results of Osmer et al.\ (1994; data adjusted to $q_o =
0$).  Metallicity variations may explain a large part of the observed
scatter. }

\clearpage

\begin{deluxetable}{lrrr}
\tablecaption{Abundance Sets\tablenotemark{a}. \label{tbl-1}}
\tablewidth{0pt}
\tablehead{
\colhead{$Z$} & \colhead{metals/H}   & \colhead{N/H}   & \colhead{He/H}} 
\startdata

0.2 & 0.2 & 0.020 & 0.942 \nl
0.5 & 0.5 & 0.125 & 0.964 \nl
1   & 1.0 & 0.500 & 1.000 \nl
2   & 2.0 & 2.000 & 1.072 \nl
5   & 5.0 & 12.500 & 1.290 \nl
10 & 10.0 & 50.000 & 1.652 \nl
 
\enddata
\tablenotetext{a} {Ratios X/H relative to solar.}
\end{deluxetable}

\clearpage

\begin{deluxetable}{llrrrrl}
\small
\tablecaption{Modelled and Observed Correlations. \label{tbl-2}}
\tablewidth{0pt}
\tablehead{
\colhead{} &  \colhead{} & \multicolumn{2}{c}{Slope $\beta$} & \multicolumn{2}{c}{Constant $c$} &  \colhead{} \\
\colhead{y} & \colhead{x}   & \colhead{Model\tablenotemark{a}}   & \colhead{Obs\tablenotemark{a}} &
\colhead{Model\tablenotemark{a}}   & \colhead{Obs\tablenotemark{a}} & \colhead{Ref\tablenotemark{b}}
} 
\startdata

log$W_\lambda$(\ion{C}{4}) & log($L_{\nu}(1550)/10^{30}$) & -0.20 & -0.25 & 1.64 & 1.73 & OPG94\nl
          &                      &       & -0.17 &      & 1.92 & KRK90\nl
          &                      &       & -0.23 &      & 2.05 & L95\nl
          &                      &       & -0.14 &      &      & ELT93\nl
          &                      &       & -0.13 &      &      & Z92\nl
          &                      &       & -0.18 &      &      & T97\nl

log$W_\lambda$(Ly$\alpha$) & log($L_{\nu}(1550)/10^{30}$) & -0.10 & -0.13 & 1.95 & 2.04 & OPG94\nl
          &                      &       & -0.08 &      & 2.12 & KRK90\nl
          &                      &       & -0.14 &      & 2.11 & L95\nl
          &                      &       & -0.06 &      &      & ELT93\nl
          &                      &       & -0.12 &      &      & T97\nl

log$W_\lambda$(\ion{O}{6}) & log($L_{\nu}(1550)/10^{30}$) &       & -0.30 &      & 1.60 & ZKD95\nl
          &                              &       & -0.15 &      & 1.43 & L95\nl
          &                              &       & -0.27 &      &      & ELT93\nl
          &                              &       & -0.30 &      &      & T97\nl

 & log($L_{\nu}(1550)/10^{30}) > -0.8$ & -0.22 && 1.43 &  & \nl    
 & log($L_{\nu}(1550)/10^{30}) < -0.8$ & -0.16 && 1.47 &   & \nl

log$W_\lambda$(\ion{N}{5}) & log($L_{\nu}(1550)/10^{30}$) &       & -0.04 &      & 1.12 & L95\nl
         &                               &       & $\sim 0$ &   & 1.36 & OPG94 \nl
         &                               &       & -0.07    &   &      & T97 \nl

 & log($L_{\nu}(1550)/10^{30}) > 1$ & -0.047 && 0.828 &   & \nl
 & log($L_{\nu}(1550)/10^{30}) < 1$ &   0 && 0.863 &  & \nl    

log$W_\lambda$(\ion{He}{2}) & log($L_{\nu}(1550)/10^{30})$ & -0.17 & -0.20 & 0.727 & 0.77 & L95\nl
         &                               &       & -0.20    &   &      & T97 \nl

log$W_\lambda$($\lambda$1400) & log($L_{\nu}(1550)/10^{30})$ & -0.13 & -0.12 & 0.985  &&     ELT93\nl
	   &                              &       & -0.25 &        & 1.26 & L95 \nl

log$W_\lambda$($\lambda$1900) & log($L_{\nu}(1550)/10^{30})$ &  &       -0.11 & & &     ELT93\nl
           &                              &  &       -0.17 & & 1.52 & L95\nl
           &                              &  &       -0.05 & &  & OPG94\nl
  &  log($L_{\nu}(1550)/10^{30}) > -0.8$ & -0.083 & &1.35 & & \nl
  &  log($L_{\nu}(1550)/10^{30}) < -0.8$ & 0 &&    1.41 \nl 
log$W_\lambda$(\ion{Mg}{2}) & log($L_{\nu}(1550)/10^{30})$ & -0.092 & -0.05 &1.45 &  &    ELT93\nl
          &                      &       &-0.08 & & & Z92 \nl

\tablebreak

log(Ly$\alpha$/\ion{C}{4}) & log($L_{\nu}(1550)/10^{30})$ & 0.10 & 0.12 & 0.314 & 0.42 &  OPG94\nl
          &                              &       & 0.09 &      &  0.25& KRK90\nl

log($\lambda$1900/\ion{C}{4}) & log($L_{\nu}(1550)/10^{30})$ &  & 0.20 &  &  &  OPG94\nl
  &  log($L_{\nu}(1550)/10^{30}) > -0.8$ & 0.12 & &-0.29 & & \nl
  &  log($L_{\nu}(1550)/10^{30}) < -0.8$ & 0.20 &&  -0.23 &&\nl

log(\ion{N}{5}/\ion{He}{2})& log(\ion{N}{5}/\ion{C}{4}) &        0.88 & 0.90 & 0.829 & 0.87 & HF93\nl

log(Ly$\alpha$/\ion{C}{4}) & $\alpha_{ox}$ & 0.89 && -1.01&&\nl

log(Ly$\alpha$/\ion{O}{6}) & $\alpha_{ox}$   &&                         0.74 &&  -0.38   &    ZKD95\nl
 &  $\alpha_{ox} > 1.4$         &        1.06 && -1.04  &&\nl 
 &  $\alpha_{ox} < 1.4$        &       0.66 && -0.478 &&\nl

\enddata

\tablenotetext{a} {Columns 3-6 give results for linear list squares
fits to $y = c + \beta x$. The model results assume $q_o = 0$.}
\tablenotetext{b} {References: \\
OPG94: Osmer, Porter, \& Green (1994). Continuum measured at 1450~\AA,
$q_o = 0.5$. \\
KRK90: Kinney, Rivolo, \& Koratkar (1990). Continuum measured locally,
$q_o = 1$.\\
L95:  Laor et al.\ (1995). Continuum measured at 1350~\AA. Limited
(1.5--3 dex) range in $L_{uv}$, $q_o = 0$.\\
ELT93: Espey, Lanzetta, \& Turnshek (1993). Continuum measured at 1450~\AA.\\
Z92: Zamorani et al.\ (1992).  Limited (2--2.8 dex) range in $L_{uv}$.
Continuum measured locally, $q_o = 1$.\\
T97: Turnshek (1997).\\
ZKD95: Zheng, Kriss, \& Davidson (1995). Continuum measured at
1550~\AA, $q_o = 1$.\\
HF93: Hamann \& Ferland (1993).}

\end{deluxetable}

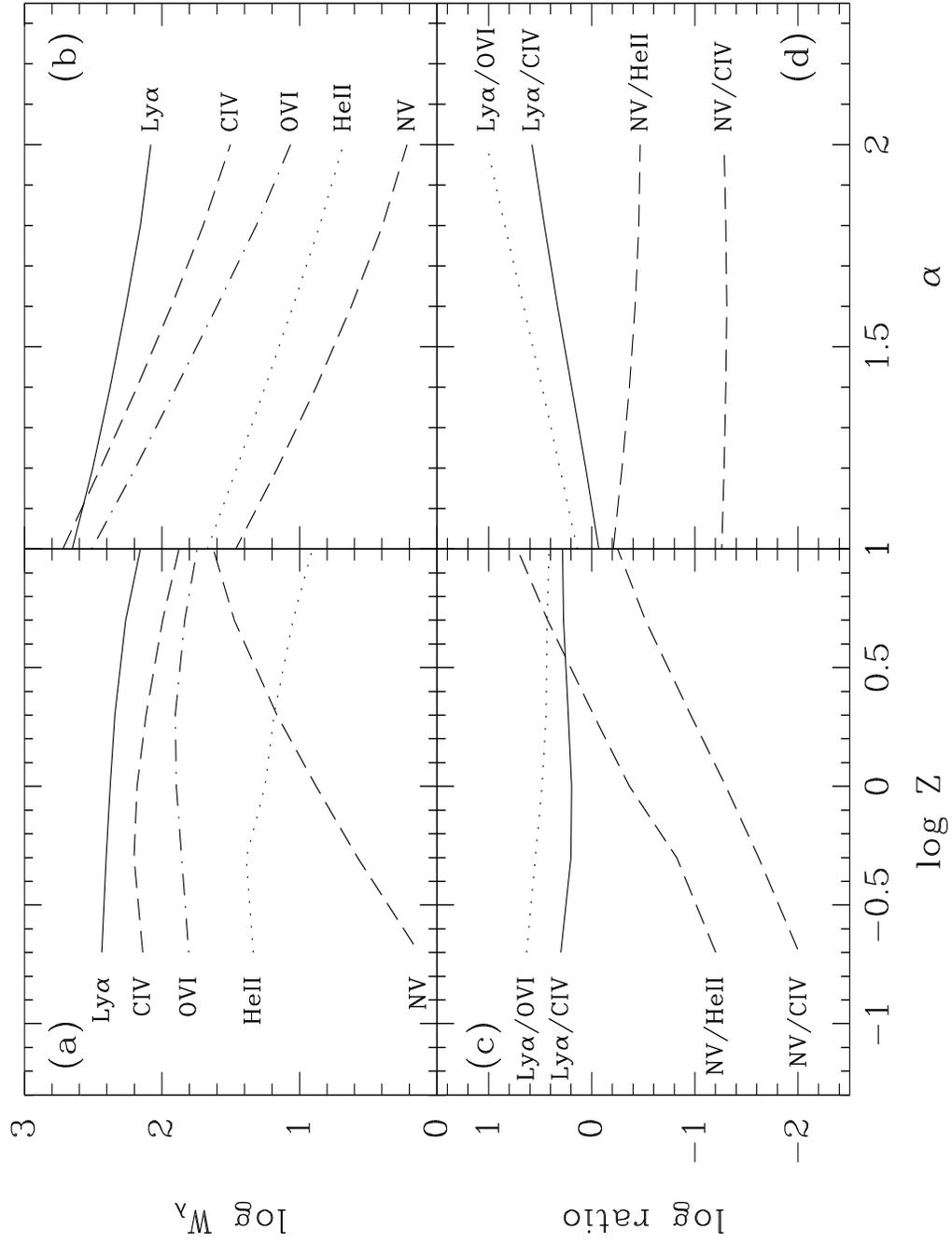
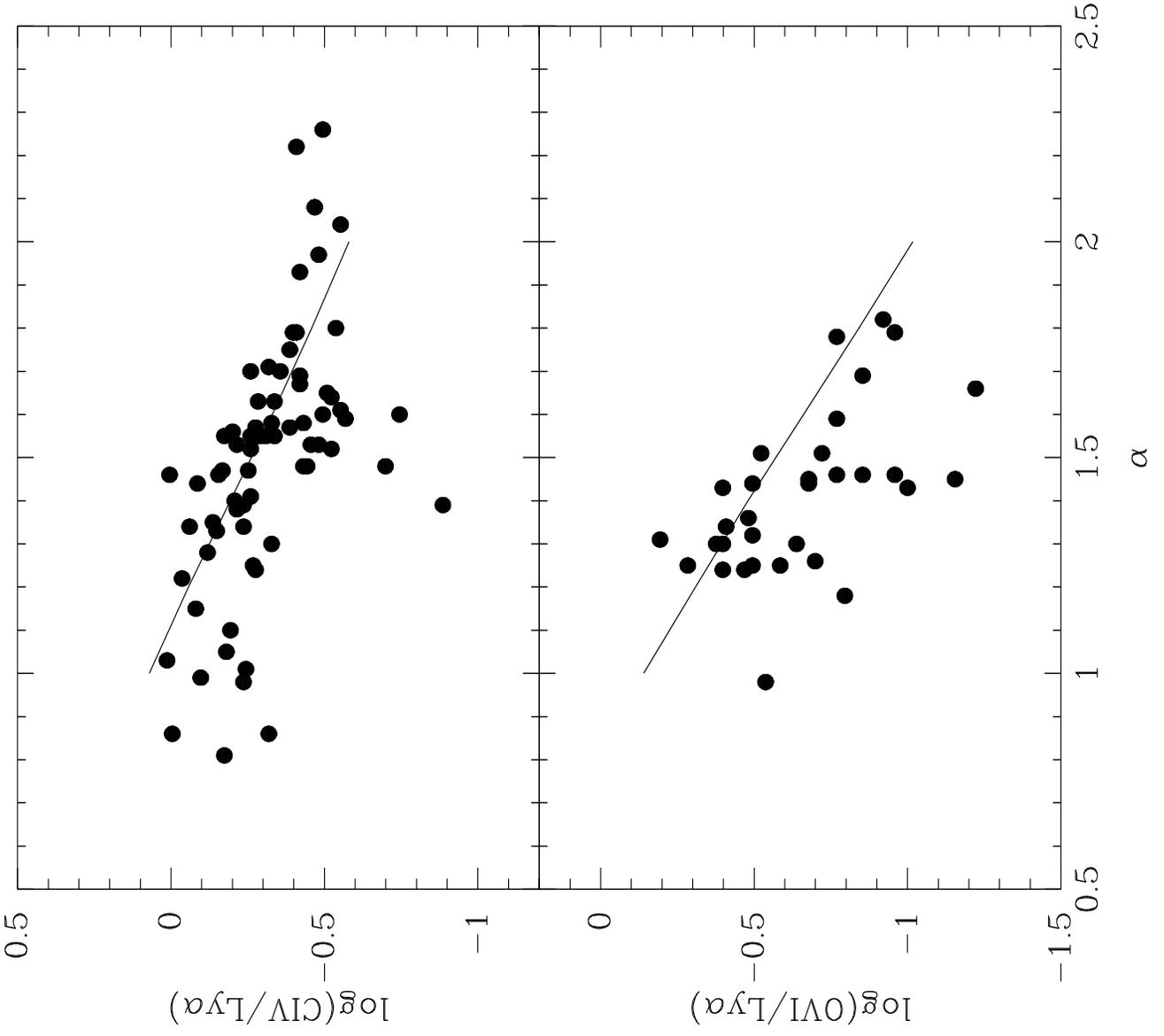
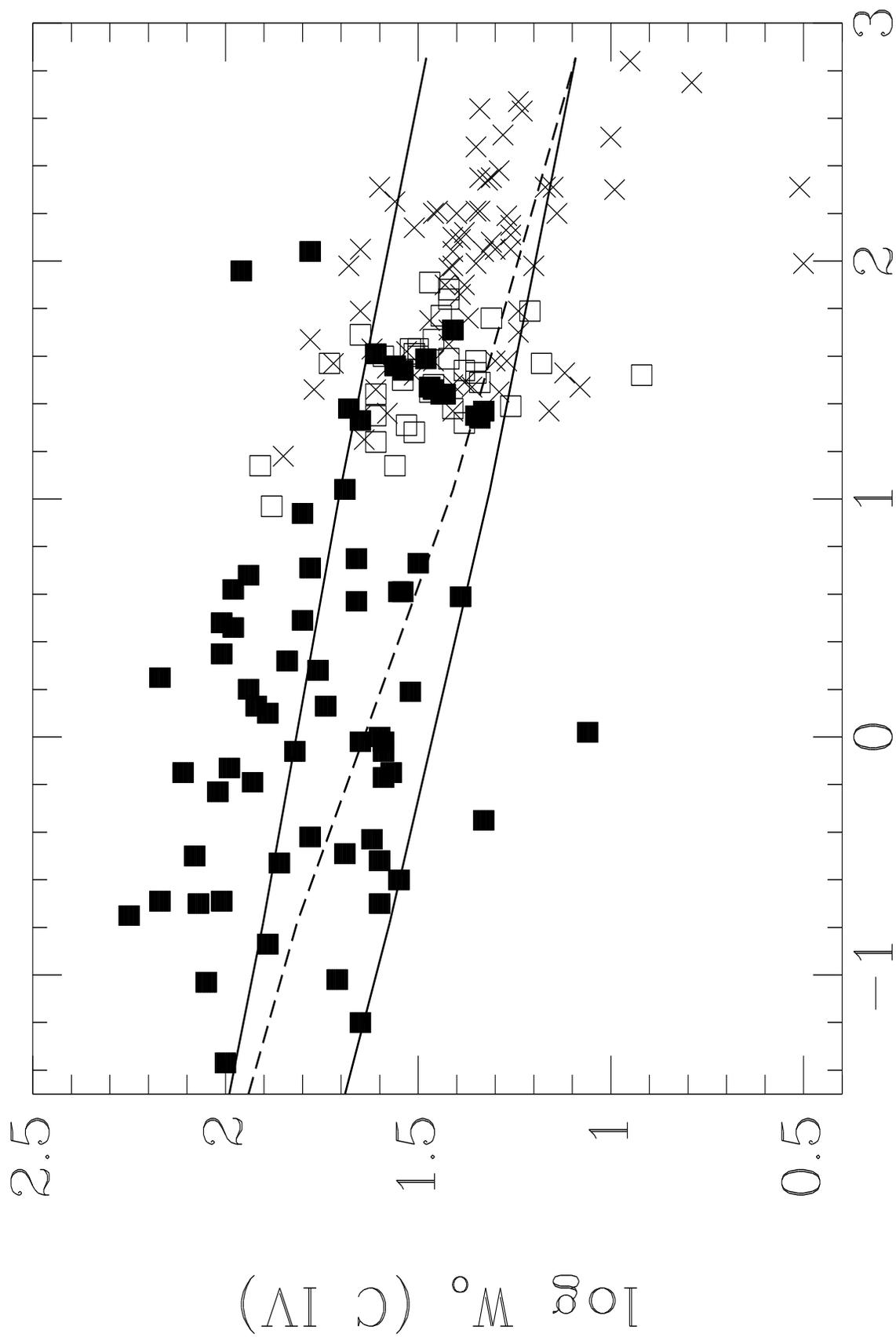
\end{document}